# Quantized conductance of Majorana zero mode in the vortex of the topological superconductor (Li$_{0.84}$Fe$_{0.16}$)OHFeSe

C. Chen[1, 2†], Q. Liu[1, 2, 3†], T. Z. Zhang[1, 2], D. Li[4], P. P. Shen[4], X. L. Dong[4], Z.-X. Zhao[4], T. Zhang[1, 2*], D. L. Feng[1, 2, 5*]

[1] State Key Laboratory of Surface Physics and Department of Physics, Fudan University, Shanghai 200433, China
[2] Collaborative Innovation Center of Advanced Microstructures, Nanjing 210093, China
[3] Science and Technology on Surface Physics and Chemistry Laboratory, Mianyang, Sichuan 621908, China
[4] National Laboratory of Superconductivity, Institute of Physics, CAS, Beijing 100190, China
[5] Hefei National Laboratory for Physical Science at Microscale, CAS Center for Excellence in Quantum Information and Quantum Physics, and Department of Physics, University of Science and Technology of China, Hefei 230026, China.

† These authors contributed equally.

**The Majorana zero mode (MZM), which manifests as an exotic neutral excitation in superconductors, is the building block of topological quantum computing. It has recently been found in the vortices of several iron-based superconductors as a zero-bias conductance peak (ZBCP) in tunneling spectroscopy. In particular, a clean and robust MZM has been observed in the cores of free vortices in (Li$_{0.84}$Fe$_{0.16}$)OHFeSe. Here using scanning tunneling spectroscopy (STS), we demonstrate that Majorana-induced resonant Andreev reflection occurs between the STM tip and this zero-bias bound state, and consequently, the conductance at zero bias is quantized as 2e$^2$/h. Our results present a hallmark signature of the MZM in the vortex of an intrinsic topological superconductor, together with its intriguing behavior.**

A peak at zero energy in tunneling spectroscopy is an important hallmark, but not sufficient proof, for identifying a MZM. For example, although the existence of a MZM is predicted at the ends of a strongly spin-orbital-coupled semiconductor nanowire in the presence of a proximate superconductor and a large Zeeman field [1-7], there are alternative interpretations for the experimental observations of zero energy peaks in these systems [8-12]. More-compelling evidence for a MZM is that the zero-bias peak possesses the quantized universal conductance, $G_0$ = 2e$^2$/h, in tunneling experiments, due to resonant Andreev reflection and the particle-hole symmetry of the MZM [13]. At zero temperature, this Majorana-induced resonant Andreev reflection (MIRAR) would give the quantized conductance regardless of the coupling strength [13]; and at finite temperature, the quantized conductance may be observed when the tunneling coupling is sufficiently strong [1,14]. The quantized conductance of a ZBCP was first observed in the tunneling spectrum of a hybrid device between superconducting aluminum and an InSb nanowire [15]. It thus strongly supports the existence of MZMs in semiconductor nanowire devices, although the ultimate proof of Majorana physics would be a demonstration of non-Abelian statistics [16-18].

The vortex core of a topological superconductor or superconducting heterostructure is another promising platform for MZMs [19-24]. For example, experimentalists have observed a ZBCP well separated from the usual off-zero-bias Caroli-de Gennes-Matricon (CdGM) states in the vortex cores of Fe(Te,Se) [25] and $(Li_{0.84}Fe_{0.16})OHFeSe$ [26]. Although a MZM seems the most likely origin of the ZBCP, other Majorana signatures such as quantized conductance have not been observed. Unlike in the more-complex nanowire systems, there is only vacuum between the tip and the surface of the topological superconductor in a scanning tunneling microscope (STM) experiment. Therefore, this provides an advantageous platform to study the properties of the MZM, such as MIRAR.

In this paper, we report a millikelvin STM study of the properties of the ZBCP in the vortex core of $(Li_{0.84}Fe_{0.16})OHFeSe$. We find that the conductance of the ZBCP indeed approaches the quantized conductance of $2e^2/h$ as the tip gradually approaches the sample surface. This provides strong evidence for the existence of a MZM in the vortex core of a topological superconductor.

High-quality single-crystalline superconducting films of $(Li_{0.84}Fe_{0.16})OHFeSe$ used here were grown on a $LaAlO_3$ substrate by a matrix-assisted hydrothermal epitaxial method, as described in Ref. [27]. The full-width at half-maximum (FWHM) of their X-ray rocking curves is 0.1~0.12 degrees, indicative of their high quality. They have a superconducting transition temperature ($T_c$) of 42 K. The measurements were conducted in a UNISOKU dilution refrigerator STM at $T$ = 20 mK. The lowest effective electron temperature ($T_{eff}$) of the system is calibrated to be 160 mK, which yields an energy resolution of $3.5k_BT_{eff}$ = 50 μV (see supplementary materials part I). The sample was cleaved at 78 K in ultrahigh vacuum with a base pressure of $5×10^{-11}$ Torr and immediately transferred into the STM module. Pt/Ir tips were used after being treated on a clean Au(111) surface. Bias voltage ($V_b$) is applied to the sample with carefully calibrated zero point (see supplementary materials part II). $dI/dV$ spectra were collected by a standard lock-in technique (with a modulation frequency $f$ = 873 Hz and a typical modulation amplitude $\Delta V$ = 0.05 mV), and by numerically differentiating the $I/V$ curve to calibrate the absolute value of tunneling conductance (see supplementary materials part III for more details).

Fig. 1(a) shows a clean region of the FeSe surface of $(Li_{0.84}Fe_{0.16})OHFeSe$ with an atomically resolved lattice (inset image). The tunneling spectra exhibit a flat-bottomed superconducting gap with a double-peak structure (Fig. 1(b)). Under a magnetic field B = 8.5T, one can find a free vortex in the defect-free region (shown in the inset of Fig. 1(c)), as reported previously [26]. Figs. 1(c) and 1(e) show the low-energy $dI/dV$ spectra taken across a free vortex core. Because of the high energy resolution here, discrete vortex states can be clearly resolved, revealing finer structures of the off-zero-bias vortex states than in our previous data taken at 0.4 K [26]. These off-zero-bias states deserve further examination in the future. Here, we focus on the sharp peak at zero energy which does not shift upon leaving the core, which was assigned as a MZM [26]. In Fig. 1(d), we show an averaged $dI/dV$ spectrum over a ±1nm region of the core center, measured at a setpoint of $V_b$ = 3 meV and $I$ = 60 pA (effective tunneling barrier resistance: $R_N$ = 5×10$^7$ Ω). A Lorentzian fit to the ZBCP gives a full-width at half-maximum (FWHM) of 0.10 meV [14], much narrower than that previously measured at T = 0.4K [26] and only slightly wider than the energy resolution here. The finite peak width arises from both tunneling broadening from the STM tip and the finite-temperature broadening, probably with some additional mechanisms such as dissipation effects.

Nevertheless, we note that the tunneling conductance of the ZBCP in Fig. 1(d) is much lower than $G_0$ ($2e^2/h$). This is due to both the low tunneling transmission or large tunneling barrier strength ($R_N \gg h/e^2$), and the finite electron temperature, which further broadens and lowers the conductance peak in the low transmission limit. Although the $T_{eff}$ of our system is only about 0.4% of $T_c$, apparently it is still not sufficient to observe the quantized conductance at such a low transmission.

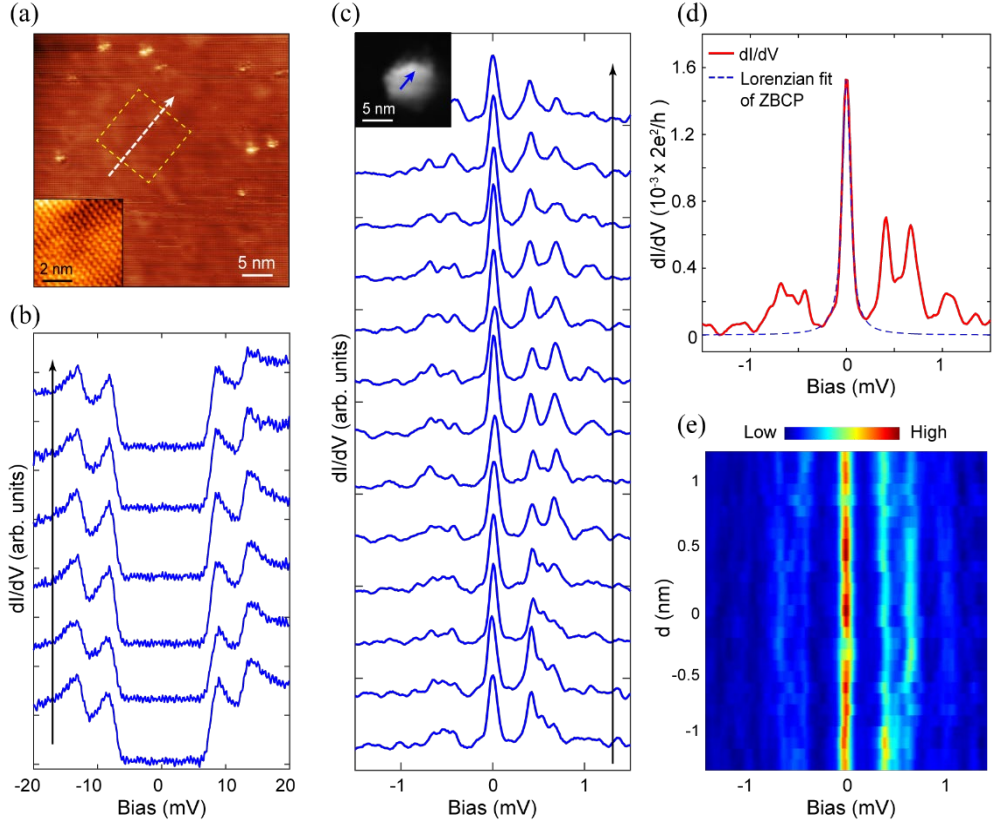

**Fig. 1.** (a) STM image of a cleaved (Li$_{0.84}$Fe$_{0.16}$)OHFeSe thin film ($V_b$ = 30 mV, $I$ = 30 pA). Inset: atomically resolved image. (b) Normalized superconducting gap spectra taken along the arrow across a defect-free area (set point: $V_b$ = 20 mV, $I$ = 80 pA). (c) Normalized $dI/dV$ spectra taken across a free vortex core (along the arrow in the inset). Inset: Zero-bias $dI/dV$ map of the square area marked in (a) under $B$ = 8.5T, which shows a free vortex core. (d) Red curve: averaged $dI/dV$ spectrum near the core center (set point: $V_b$ = 3 mV, $I$ = 60 pA). The tunneling conductance is calibrated by scaling to the numerical differential of the I/V curve (see SM part III). Blue dashed curve: Lorentzian fit to the ZBCP, with a FWHM = 0.10 meV. (e) Spatial dependence of the $dI/dV$ spectra in panel (c), shown in a false-color plot.

To enhance the tunneling transmission, we gradually reduce the distance between the STM tip and the sample surface. This is achieved by increasing the setpoint tunneling current ($I_{set}$) at fixed bias voltage ($V_b$). We then define $G_N = I_{set}/V_b$ which reflects the transmissivity of the tunneling barrier. In Fig. 2, we show the $dI/dV$ spectra taken at the cores of two free vortices while incrementally reducing the tip-sample distance, where the starting $I_{set}$ is already much higher than that for data in Fig. 1. For Vortex 1 (Fig. 2(a)), as $G_N$ increases the conductance of the ZBCP first increases rapidly, then starts to saturate around 0.9 $G_0$. The off-zero-bias peaks are greatly enhanced as well. Figs. 2(b)-2(e) show four representative spectra at different $G_N$: at low $G_N$ (Figs. 2(b) and 2(c)), the spectral line shape evolves slowly with increased $G_N$, and the ZBCP remains sharp. However, starting at $G_N = 0.36$ $G_0$ (Fig. 2(d)), the off-zero-bias states grow quickly in the spectrum, which causes broadening of the central peak. In the high-transmission cases such as that shown in Fig. 2(e), the two side peaks off zero bias become higher than the zero-bias peak. However, it is notable that conductance at zero-bias remains more or less fixed, within the experimental uncertainty.

Fig. 2(f) shows the data of another vortex (Vortex 2), where the ZBCP is weaker than that shown in Fig. 2(a). In this case, we managed to further increase $G_N$ beyond $G_0$, and found that although the side peaks can exceed $G_0$ at sufficiently large $G_N$, the zero-bias conductance saturates around $G_0$ over an extended $G_N$ range. This is a remarkable feature which points to the quantization of zero-bias conductance. Figs. 2(g)-2(l) show six representative spectra at different $G_N$'s. At low $G_N$, only the ZBCP is pronounced (Fig. 2(g)). With increased $G_N$, two side peaks show up and the feature at zero bias is significantly broadened (Figs. 2(h)-2(j)). Then the spectrum turns into a double-peak structure (Figs. 2(k)-2(l))) which keeps the zero-bias conductance around $G_0$. Such a broadening observed in both vortices indicates strong tunneling broadening due to strong coupling between the tip and the sample, which enables the observation of quantized conductance [1]. Eventually, as the tip further approaches the sample surface, the two off-zero-bias peaks are further enhanced, and they merge and overwhelm the zero-bias peak (magenta curves in Fig. 2(m)). As a result, the zero-bias conductance seems to suddenly exceed $G_0$.

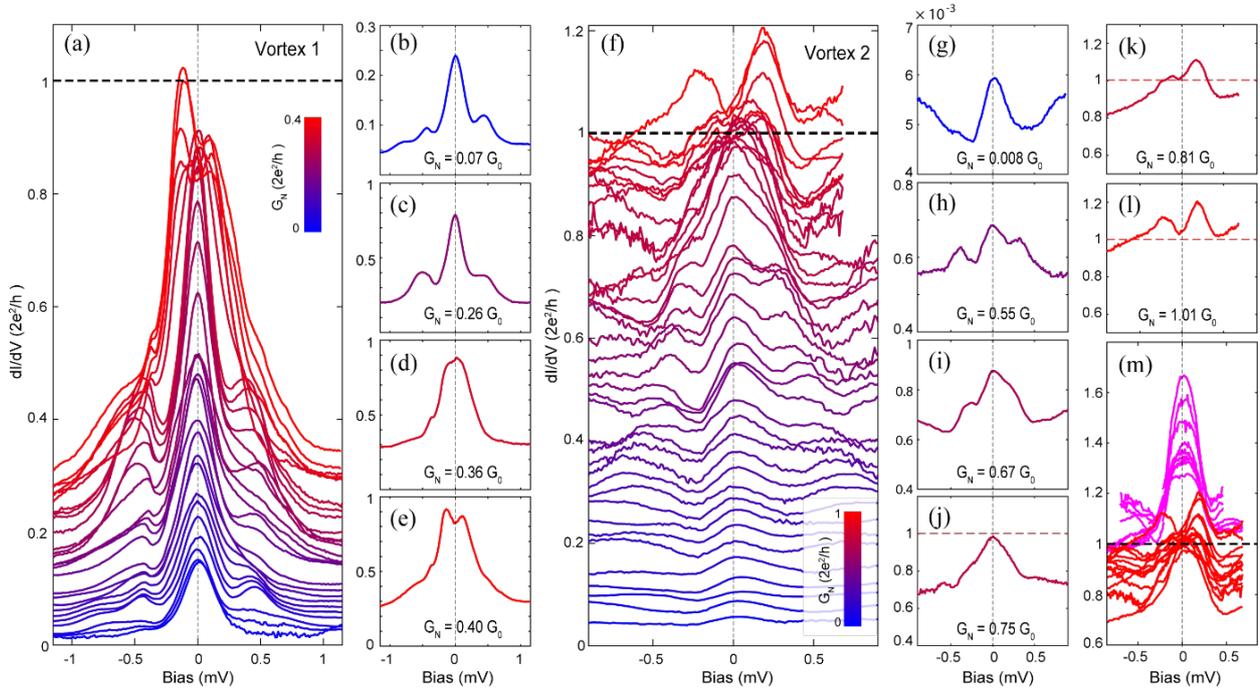

**Fig. 2**. (a) Evolution of the *dI/dV* spectra as a function of increased tunneling transmission (reduced tunneling barrier strength) reflected by $G_N = I_{set}/V_b$ ($V_b = -1.7$mV), for Vortex 1. The tunneling conductance is calibrated in a similar way to that described in SM part III. (b-e) Selected dI/dV spectra taken at different $G_N$ for Vortex 1. (f) Evolution of the *dI/dV* spectra as function of $G_N$ for Vortex 2 ($V_b$

= -0.9mV). (g-l) Selected $dI/dV$ spectra taken at different $G_N$ for Vortex 2. (m) Evolution of the $dI/dV$ spectra for $G_N$'s in the range of (0.8 $G_0$, 1.2 $G_0$) for Vortex 2, where a zero-bias peak with conductance exceeding $G_0$ is observed. The spectra in (f-m) are obtained by directly differentiating the $I/V$ curve.

We note that to enhance $G_N$ or tunneling transmission, we have gradually increased $I_{set}$ up to 80 nA. To examine whether such a large tunneling current may affect the superconducting properties, we checked the superconducting spectrum outside the vortex core, as shown in Fig. S4 of the supplementary materials. The superconducting gap size is slightly reduced at high current, but the overall line shape remains the same.

Fig. 3 summarizes the zero-bias conductance as a function of $G_N$ from two extensive sets of $dI/dV$ spectra, including those in Fig. 2. The zero-bias conductance of Vortex 1 gradually increases with increased transmission (Fig. 3(a)), and an extrapolation suggests that it may approach $G_0$ with further enhanced transmission. The data for Vortex 2 extends much further, and can be separated to three regimes. In the first regime, the zero-bias conductance increases approximately linearly with $G_N$ (black dashed line); in the second regime, it exhibits saturation around $G_0$ (red dashed line); and in the third regime, it exhibits a steep enhancement (magenta dashed line).

The relatively flat zero-bias conductance around $G_0$ in the second regime represents compelling evidence that the ZBCP indeed possesses the quantized conductance of $2e^2/h$, which renders strong support to an MZM origin. Since this regime extends between $G_N$=0.75 $G_0$ and $G_N$=1.0 $G_0$, it indicates that the quantized conductance of the MZM can be fully detected even when the transmission of the off-zero-bias states is less than unity. That is, the resonant tunneling of the MZM occurs for $G_N$ > 0.75 $G_0$ at $T_{eff}$ = 160 mK, which demonstrates nicely the MIRAR effect at finite temperature and in the strong coupling regime between the tip and vortex. We note that although the 160 mK electronic temperature here is just 0.4% of $T_c$, it is about 8% of the mini-gap between the ZBCP and the first excited vortex state, which is about 0.2 meV (Fig. 2(f)). Therefore, strong coupling and high transmission are needed, to ensure that the tunneling broadening of the ZBCP due to the STM tip is much larger than the finite temperature broadening. Only in this limit can the MZM-induced ZBCP approach $2e^2/h$ asymptotically [1,14].

The steep enhancement in the third regime is likely due to topological trivial states. As sketched in the inset of Fig. 3(b), when the tip is sufficiently close to the sample, besides the tunneling between the tip and the MZM at the vortex center, the tunneling to the region surrounding the core center will be enhanced. Therefore, some off-zero-bias vortex states located farther from the center will start to contribute more to the $dI/dV$ spectrum [28]. In Figs. 2(f)-2(l), one can see that the off-zero-bias peaks are strong in the high transmission regime. Because of their finite widths due to tunneling broadening, dissipation, etc., these peaks may contribute additional conductance at zero bias. These effects could be responsible for the steep enhancement above $G_0$. The third regime is thus actually a multi-channel tunneling regime. Furthermore, because different vortex states may have different radial distributions, their contributions to the STS spectrum may vary dramatically as the tip approaches the sample. Consequently, the spectral lineshape may be greatly affected, as shown in Fig. 2.

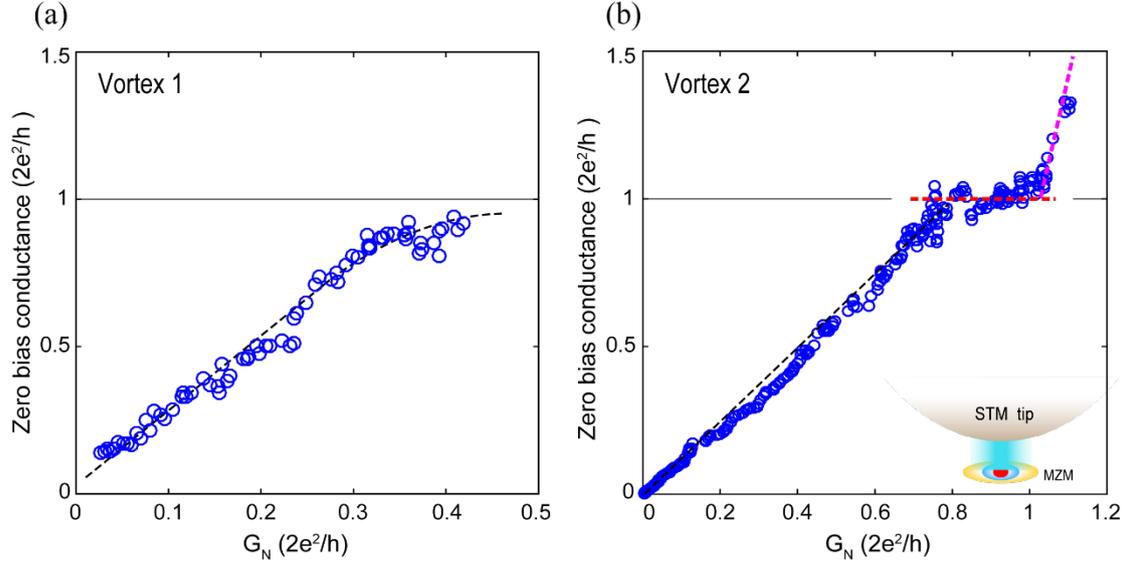

**Fig. 3.** Summary of the zero-bias conductance as a function of $G_N$ for (a) Vortex 1, and (b) Vortex 2. The inset of panel b depicts the tunneling between the STM tip and the vortex states, where the MZM (red) is mostly located in the center of vortex core, while some other off-zero-bias vortex states (cyan and yellow) surround it. Due to the finite size of the tip, the tunneling between the tip and these vortex states is enhanced when the tip is sufficiently close to the sample, even when the sharpest point of the tip is at the center.

To conclude, we have shown that the zero-bias mode in the vortex core of the $(Li_{0.84}Fe_{0.16})OHFeSe$ superconductor exhibits quantized conductance of $2e^2/h$, further supporting the existence of Majorana zero modes in the vortices of $(Li_{0.84}Fe_{0.16})OHFeSe$ by illustrating its characteristic hallmark. Moreover, the fact that the quantized conductance is observed before the transmission of the off-zero-bias states reaches unity demonstrates Majorana-induced resonant Andreev reflection, a fascinating effect, at finite temperature. Our results facilitate a further understanding of Majorana zero mode and suggest that the iron selenide superconductors as a promising platform for the investigation of exotic Majorana physics and the development of topological quantum computing.

We thank Dr. Chun-Xiao Liu, Prof. Qianghua Wang, Prof. Yajun Yan and Dr. Darren Peets for helpful discussions. This work is supported by the National Natural Science Foundation of China, National Key R&D Program of the MOST of China (Grant No. 2016YFA0300200, 2017YFA0303004, and 2017YFA0303003), and the Key Research of Frontier Sciences of CAS under Grant No QYZDY-SSW-SLH001.

* tzhang18@fudan.edu.cn,
* dlfeng@fudan.edu.cn

# Supplementary Materials

## I. Calibration of STM energy resolution.

Due to electrical noise and RF radiation, the effective electron temperature ($T_{eff}$) of a low-T STM is usually higher than the thermometer reading. Proper electronic shielding and RF filtering are necessary to reduce $T_{eff}$. We calibrated the $T_{eff}$ of our dilution refrigerator STM at $T = 20$ mK by measuring the superconducting gap of an Al film on Si(111). Fig. S1 shows the measured gap spectrum – a standard Maki BCS fitting [1] yields $\Delta = 0.19$ meV, $\zeta = 0.02$ meV and $T_{eff} = 160$ mK.

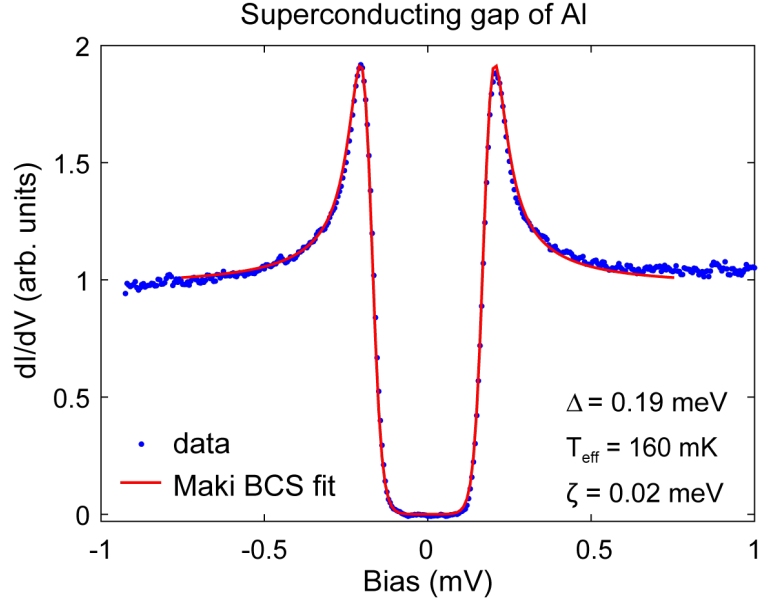

**Figure S1**: Superconducting gap spectrum of 20 ML Al film on Si(111) ($V_b = 1$mV, $I = 100$ pA, $\Delta V = 20$ µV). Red curve is the Maki BCS fit. Fitting parameters: $\Delta = 0.19$ meV, $\zeta = 0.02$ meV, $T_{eff} = 160$ mK.

## II. Calibration of STM bias offset.

The STM bias voltage eventually applied to the sample usually has a small offset. Such an offset can be calibrated by measuring $I$-$V$ curves at different setpoints ($I_{set}$), because all the $I$-$V$ curves should intersect at a single point where $V = 0$ and $I = 0$. Fig. S2 shows such a calibration performed on the vortex core state measurement, which yields a bias offset of 0.32 meV. All the tunneling spectra presented in this paper are calibrated in the same way.

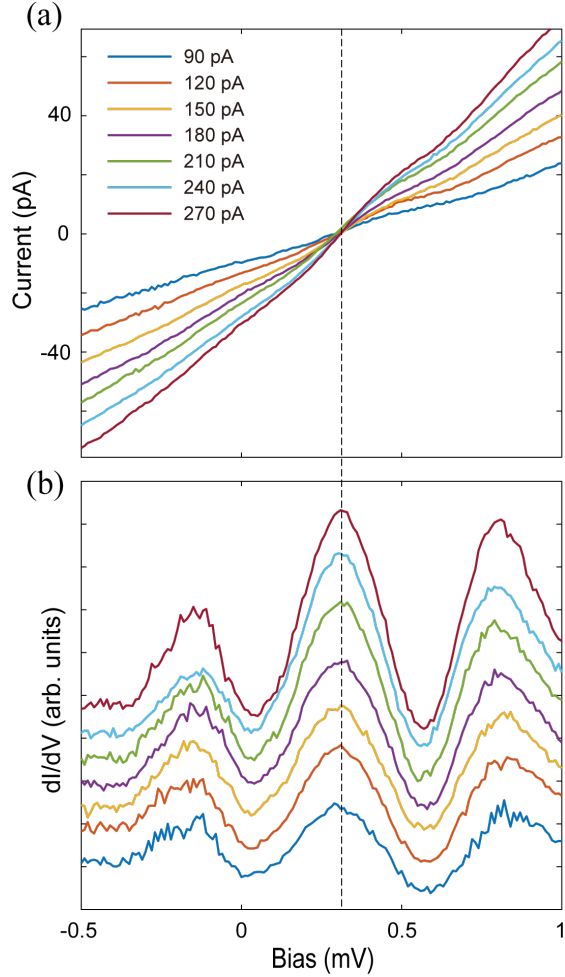

**Figure S2**: (a) Raw *I*/*V* curve measured at different setpoint ($I_{set}$). (b) simultaneously obtained raw *dI*/*dV* curve.

## III. Calibration of the tunneling conductance value of the *dI*/*dV* spectra

The absolute value of tunneling conductance is very important for this work. In a standard lock-in measurement, the result is proportional to the differential tunneling conductance (*dI*/*dV*) but does not have an absolute value. In fact, the absolute *dI*/*dV* can be directly calculated from numerical differentiation of the *I*/*V* curve, but usually has low signal/noise ratio. In Fig. S3(a) we show the *I*/*V* curve taken simultaneously with the STS shown in Fig. 1(d), and its numerical differential with absolute conductance. Then we scale the lock-in measured *dI*/*dV* to match the line shape of numerical differential *dI*/*dV*, as shown in Fig. S3(b), thereby calibrating the lock-in-detected *dI*/*dV*. The spectra shown in Fig. 2(a) are calibrated in this way. On the other hand, once the tunneling barrier is significantly reduced (tunneling current is high), the signal/noise of the numerical differential becomes high enough to use directly. In this case we use the numerical differential *dI*/*dV* to calculate the conductance, as shown in Fig. 2(f).

We also note that the wire resistance of the bias voltage/tunneling current lines of the STM, which is usually hundreds of ohms, could induce non-negligible error in such low-tunneling-barrier measurements, since they are in series with the tunneling junction. Therefore, we carefully measured the wire resistance *in-situ* by shorting the tip to the sample after the whole experiment, and exclude this contribution from the total resistance of the tunneling loop.

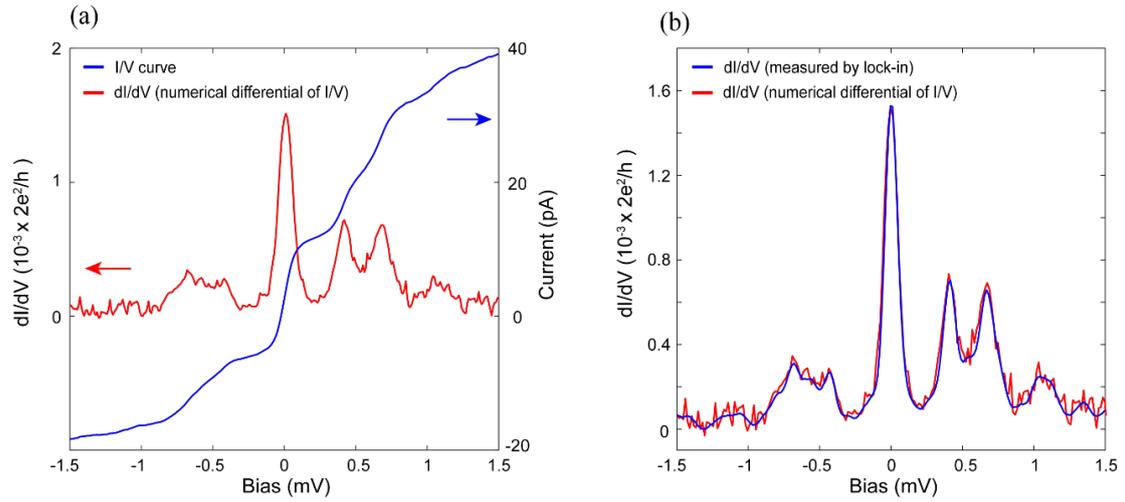

**Figure S3**: Calibration of the tunneling conductance value. (a) *I/V* curve and its numerical differential taken simultaneously with the STS shown in Fig. 1(d). (b) Comparison of the lock-in measured *dI/dV* and numerical differential of the *I/V* curve.

## IV. Measurement of the superconducting gap at high tunneling current.

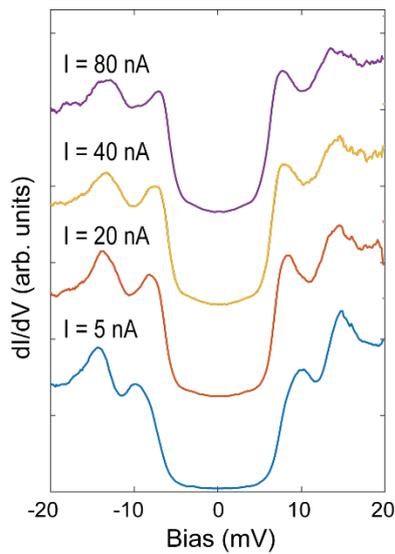

**Figure S4**: Normalized *dI/dV* spectra measured on a superconducting region outside the vortices, as a function of the tunneling current measured at 20 mV.

**References**
[1] Maki K 1964 *Prog. Theor. Phys.* **31** 945